\documentclass[
	aps,
	prl, 
	twocolumn, 
	a4paper, 
	10pt, 
	amsmath,
	amssymb, 
	preprintnumbers,  
	tightenlines, 
	notitlepage,
	floatfix,
	longbibliography
	]{revtex4-1}\pdfoutput=1
\usepackage[utf8]{inputenc}
\usepackage[english]{babel}
\usepackage{bbm}
\usepackage{graphicx}
\usepackage{hyperref}
\usepackage{wasysym}
\usepackage{amsthm}
\usepackage{dcolumn}
	\newcolumntype{.}{D{.}{.}{13}}
	\newcolumntype{d}[1]{D{.}{.}{#1}}
\usepackage{color}	
\usepackage[all,cmtip]{xy}
\usepackage{tensor}

\graphicspath{{pics/}}

\allowdisplaybreaks[1]

\newcommand{\abs}[1]{\lvert#1\rvert}			
\newcommand{\avg}[1]{\left\langle #1 \right\rangle }		

\newcommand{\cbB}[1]{\Big\{#1\Big\} }			
\newcommand{\hh}[1]{\left(#1\right) }			

\renewcommand{\d}[2]{\frac{\operatorname{d}\!#1}{\operatorname{d}\!#2}}					

\newcommand{\mr}{\eta}

\DeclareMathOperator{\bigO}{\mathcal{O}}					

\newcommand{\prlsection}[1]{\paragraph*{#1.\textendash{}}}

\begin{document}

\title{Self-force corrections to the periapsis advance around a spinning black hole}

\author{Maarten \surname{van de Meent}}
\email{M.vandeMeent@soton.ac.uk}
\affiliation{Mathematical Sciences, University of Southampton, Southampton, SO17 1BJ, United Kingdom}

\date{\today}
\begin{abstract}
The linear in mass ratio correction to the periapsis advance of equatorial nearly circular orbits around a spinning black hole is calculated for the first time and to very high precision, providing a key benchmark for different approaches modelling spinning binaries. The high precision of the calculation is leveraged to discriminate between two recent incompatible derivations of the 4PN equations of motion. Finally, the limit of the periapsis advance near the innermost stable orbit (ISCO) allows determination of the ISCO shift, validating previous calculations using the first law of binary mechanics. Calculation of the ISCO shift is further extended into the near extremal regime (with spins up to $1-a=10^{-20}$), revealing new unexpected phenomenology. In particular, we find that the shift of the ISCO does not have a well-defined extremal limit, but instead continues to oscillate.
\end{abstract} 

\maketitle
\setlength{\parindent}{0pt} 
\setlength{\parskip}{6pt}

\prlsection{Introduction}
The periapsis advance has been one of the key observables used to benchmark theoretical models of binary dynamics, comparing both between models and to observations. The anomalous rate of Mercury's perihelion advance had been a great source of mystery when, in 1915, it was explained by Einstein's theory of general relativity, thereby providing the first successful test of the new theory \cite{Einstein:1915}. Einstein's calculation was done using a weak field approximation appropriate for Mercury's orbit. Nowadays, the advent of gravitational wave astronomy requires the modelling of highly relativistic binary systems composed of compact objects such as black holes and neutron stars, where the periapsis advance can be multiple radians per orbit. Unfortunately, the non-linear Einstein equations do not allow for analytic solutions of the binary dynamics. Instead we have to rely on various approximation schemes, including post-Newtonian (PN) expansions \cite{Blanchet:2013haa}, expansion in the mass-ratio \cite{Poisson:2011nh}, effective one body (EOB) models \cite{Buonanno:1998gg}, and numerical discretization of the nonlinear equation in numerical relativity (NR) \cite{Centrella:2010mx}.

Calculations in each scheme are highly complex and have their own domain of validity. It is therefore of key importance to be able to compare results between different approximation schemes for both validation of the calculations and establishing where the various approximations break down. This requires the calculation of coordinate invariant observables. The periapsis advance of nearly circular orbits and the shift of the innermost stable circular orbit (ISCO) are two such observables. In the case of non-spinning binaries, these have previously been calculated and compared in Ref. \cite{LeTiec:2011bk} using a variety of different methods including self-force, PN, EOB, and NR.

This Letter focuses on the dynamics of spinning binary black holes with aligned spin and orbital angular momentum (a.k.a. ``equatorial'' binaries) in the limit that one black hole is much more massive than the other. The need for modelling such systems has recently been highlighted by the observation of GW150914 \cite{Abbott:2016blz}, which hinted at the existence of a population of massive {30-50~$M_{\astrosun}$} black holes, thereby raising the possibility of observing mergers with relatively low mass-ratios $\sim 1:20$; a regime where the faithfulness of the current template banks may be questioned. Furthermore, binary mergers with even more extreme mass-ratios ($1:10^5$) form a key science target for future space based gravitational wave observatories such as LISA scheduled for launch in the early 2030s.

Ignoring radiation reaction, equatorial binary systems are characterized by two frequencies: the radial frequency $\Omega_r$ and the averaged azimuthal motion $\Omega_\phi$, defined by
\begin{equation}
\Omega_r:=\frac{2\pi}{P_r},\quad \Omega_\phi := \avg{\d{\phi}{t}}_t,
\end{equation}
where $P_r$ is the period between two successive periapsis passes observed by an asymptotic inertial observer in the center-of-mass frame, and $\avg{\cdot}_t$ denotes averaging (with respect to the asymptotic observer's time $t$) over one radial period $P_r$. These frequencies are coordinate invariant observables that can serve to identify a particular orbit. In the limit of circular equatorial orbits, the relation between $W:=\Omega_r^2/\Omega_\phi^2$ and $\Omega_\phi$ is a coordinate invariant that measures the periapsis advance and can be used to compare between different calculation schemes. More precisely, $W$ is invariant only under a restricted class of coordinate transformations and is therefore referred to as a ``quasi-invariant'' (see \cite{gaugecompletion}).

This Letter provides the first direct numerical calculations of the (exact) linear in mass-ratio correction to $W$ around a spinning black hole using the gravitational self-force (GSF) formalism (the non-spinning case was first presented in \cite{Barack:2010ny}). These are compared to previous estimates using PN and NR calculations. The high numerical precision of our calculations further allows us to discriminate between two recent (and apparently incompatible) calculations of 4PN equations of motion for non-spinning binaries \cite{Jaranowski:2013lca,Damour:2014jta,Jaranowski:2015lha,Bernard:2015njp}. Moreover, calculation of the periapsis shift at the ISCO allows us to calculate the shift of the ISCO in a fully dynamical way independent of any external assumptions. We compare this result to earlier calculations of the ISCO shift \cite{Isoyama:2014mja} using the first law of binary mechanics and the Hamiltonian GSF framework. Finally, we study the limit of the ISCO shift for extremal spins, revealing unexpected new phenomenology.

\prlsection{Formalism}
The basic scenario studied in this Letter consists of a pair of black holes with masses $m_1$ and $m_2$, where the mass-ratio $\eta:=m_2/m_1$ is very small. The primary black hole is allowed to have a spin $\abs{a}=\abs{s_1}/m_1^2<1$ aligned with the orbital angular momentum (negative values of $a$ indicate spin anti-aligned with the orbital angular momentum). We further use geometrized units such that $G=c=1$.

The linear-in-mass ratio correction to $W$ in the circular orbit limit for non-spinning binaries was first studied in \cite{Damour:2009sm,Barack:2010ny}. Their analysis can straightforwardly be extended to spinning binaries. Following \cite{Damour:2009sm,Barack:2010ny}, the linear in mass-ratio correction to the periapsis advance is defined through
\begin{equation}\label{eq:rhodef}
W(\mr;a,\tilde\Omega_\phi) = W(0;a,\tilde\Omega_\phi) + \mr \rho(a,\tilde\Omega_\phi) +\bigO(\mr^2),
\end{equation}
where $\tilde\Omega_\phi = (m_1+m_2)\Omega_\phi$. The background value is given by
\begin{equation}
W(0;a,\tilde\Omega_\phi) = 1 - 6 x + 8 a x^{3/2} - 3 a^2 x^2,
\end{equation}
where $x := (\tilde\Omega_\phi^{-1}-a)^{-2/3}$, such that at leading order in $\eta$ we have the approximation $x\approx m_1/r$ with $r$ the radius of the background orbit.

Generalizing the derivation of \cite{Damour:2009sm,Barack:2010ny} to Kerr spacetime (utilizing key parts of the analysis done in \cite{Warburton:2011hp,Barack:2010ny}) we obtain an expression for $\rho$ in terms of the gravitational self-force $F^\mu$ on slightly eccentric orbits,
\begin{equation}
\begin{split}
\rho(a,\tilde\Omega_\phi) =\lim_{e\to 0}2&\frac{1-3x +2 a x^{3/2}}{x}\cbB{\frac{
1
}{
2x
}F^r_1
\\
&-\frac{
	1-3x +2 a x^{3/2}+a^2 x^2
	}{
		 \sqrt{1-6x+8a x^{3/2}-3 a^2 x^2}
	}F_\phi^1\\
&-\frac{
	ax^{1/2}-3x+ax^{3/2}+a^2 x^2
}{
	\sqrt{1-6x+8a x^{3/2}-3 a^2 x^2}
}a F_t^1\\
&-\frac{
	1-x(1+4ax^{1/2}-4a^2 x^2)
}{
x(1-2x+a^2x^2)
}F^r_0
}\\
&+2x\hh{1+ax^{3/2}}\hh{1-a x^{1/2}}^2,
\end{split}
\end{equation}
where $e\ll 1$ is the eccentricity, and
\begin{equation}
\begin{aligned}
F^r_0 &:=\avg{F^r}_t, &
F^r_1 &:= \frac{2}{ e}\avg{\cos(\Omega_r t) F^r}_t, \\
F_\phi^1 &:=\frac{2}{e}\avg{\sin(\Omega_r t) F_\phi}_t, &
F_t^1 &:=\frac{2}{e}\avg{\sin(\Omega_r t) F_t}_t,
\end{aligned}
\end{equation}
where $F_\mu$ and $F^\mu$ are co-/contra-variant components of the gravitational self-force along the orbit (see \cite{vandeMeent:2016pee} for a brief review and conventions).

Another key coordinate invariant observable is the shift of the ISCO. This has previously been calculated for spinning binaries in \cite{Isoyama:2014mja}, which used a Hamiltonian formulation of the conservative self-force dynamics and the first law of binary of binary mechanics \cite{LeTiec:2011ab,Friedman:2001pf,Blanchet:2012at,Tiec:2013kua} to extract the ISCO shift from data for the redshift invariant on circular orbits. It would be desirable to do an independent calculation of the ISCO shift from the self-forced dynamics, as has previously been done for non-spinning binaries \cite{Barack:2009ey}.

In \cite{Damour:2009sm,Barack:2010ny}, it was observed (for non-spinning binaries) that since the ISCO is defined by the condition that $\Omega_r=0$, calculating $\rho$ at the ISCO was equivalent to obtaining the ISCO shift. This remains true for spinning binaries. If, following \cite{Damour:2009sm,Isoyama:2014mja}, we define
\begin{equation}
(1+\mr)\Omega_{\phi}^{ISCO} :=  \breve\Omega_{\phi}^{ISCO}\hh{1+\mr C_\Omega(a)+\bigO(\mr^2)},
\end{equation}
where $\breve\Omega_{\phi}^{ISCO}$ is the ISCO frequency in the background spacetime, then observing that at the ISCO $W=0$, Eq.~\eqref{eq:rhodef} can be solved for $C_\Omega$ to obtain
\begin{equation}\label{eq:GSFisco}
C_\Omega(a) = \frac{
\rho(a,\breve\Omega_{\phi}^{ISCO})
}{
4 x_{ISCO} (1+a x^{3/2}_{ISCO})(1-ax^{1/2}_{ISCO})^2
},
\end{equation}
which generalizes Eq. (24) in \cite{Barack:2010ny}. This provides an alternative method to \cite{Isoyama:2014mja} for calculating $C_\Omega$, which has the advantage that it obtains the ISCO shift directly from the orbital dynamics rather than first passing through a Hamiltonian formulation (and any accompanying assumptions).

\prlsection{Numerical Method}
The redshift and gravitational self-force are obtained numerically using the methods described previously in \cite{vandeMeent:2015lxa} and \cite{vandeMeent:2016pee}, respectively. In these, the local metric is reconstructed in a radiation gauge from a solution of the Teukolsky equation using the formalism of Chrzanowski, Cohen, and Kegeles \cite{Chrzanowski:1975wv,Cohen:1974cm,Kegeles:1979an}. The solutions of the Teukolsky equation are obtained to very high precision using a numerical implementation \cite{Meent:2015a} of the semi-analytic method of Mano, Suzuki, and Takasugi (MST)\cite{Mano:1996gn,Mano:1996vt}. This procedure recovers the local metric only up to perturbations to the mass and angular momentum of the spacetime, which are recovered using the analysis of \cite{Merlin:2016boc}.

The gravitational self-force is obtained from the radiation gauge metric perturbation using the `no string gauge' prescription of \cite{Pound:2013faa}. In this prescription, the gauge has a discontinuity on a hypersurface containing the particle worldline. Consequently, the local time coordinate $t$ is not directly related to the time of an asymptotic inertial observer, which makes the gauge unsuitable for calculating the quasi-invariants sought in this Letter. A general procedure for calculating quasi-invariants in this class on arbitrary orbits using the `no string' prescription will be given in \cite{gaugecompletion}. The gist of that analysis is that it is sufficient to determine the stationary axisymmetric part of the gauge perturbation inside the particle orbit, which can be fixed uniquely by generalizing the procedure of \cite{Merlin:2016boc} to require the continuity of all metric components in a specified reference gauge.

\prlsection{Results}
\begin{figure}[tb]
\includegraphics[width=\columnwidth]{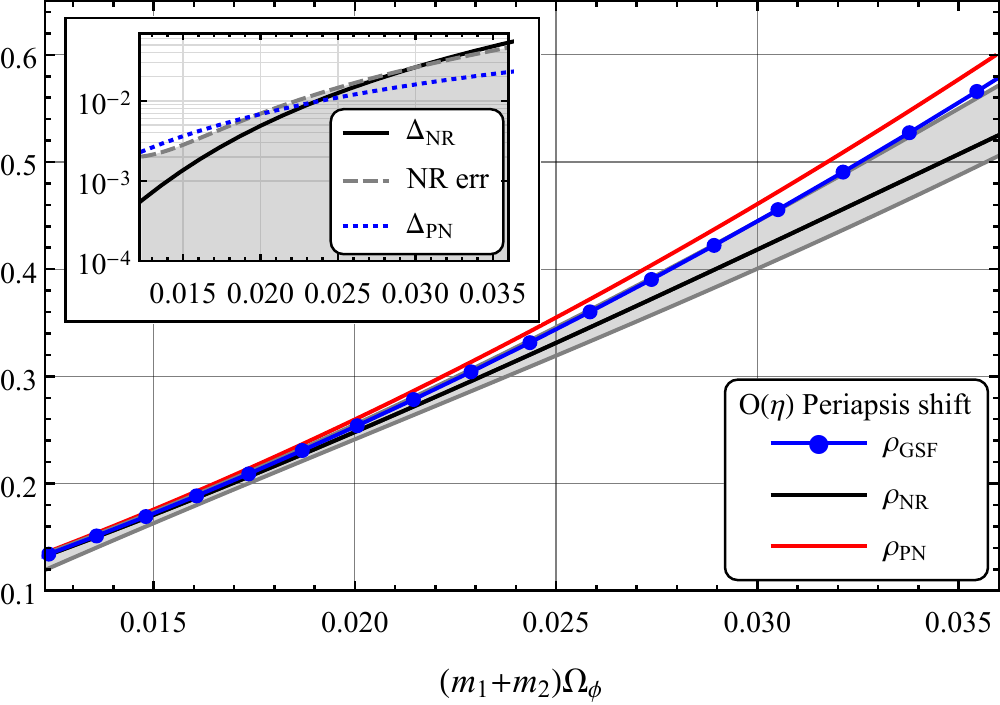}
\caption{Comparison of our (exact) numerical calculation of the linear in mass ratio correction to the periapsis advance, $\rho_{\mathrm{GSF}}$, to previous NR $\rho_{\mathrm{NR}}$ and PN $\rho_{\mathrm{PN}}$ estimates at $a=-0.5$ provided in \cite{Tiec:2013twa}. The inset shows the differences $\Delta_{\mathrm{NR}}=\abs{\rho_{\mathrm{NR}}-\rho_{\mathrm{GSF}}}$ and $\Delta_{\mathrm{PN}}=\abs{\rho_{\mathrm{PN}}-\rho_{\mathrm{GSF}}}$ on a semi-Log-scale. The shade region indicates the error on the NR estimate.
}
\label{fig:PAcompare}
\end{figure}
We have calculated the the periapsis shift $\rho(a,\tilde\Omega_\phi)$ over a range of background orbits with spin $a$ ranging from $-0.9$ to $0.9$ and $\Omega_\phi$ ranging from $10^{-3}$ to $\Omega_\phi^{ISCO}$ on a logarithmic scale. The full numerical results are available as Supplement Material~\cite{SD}.

In \cite{Tiec:2013twa} Le Tiec et al. provided an estimation of the linear in mass ratio correction to the periapsis advance in two ways: (i) using an (almost) 3.5 PN approximation, and (ii) by fitting to a series of NR simulations at $a=-0.5$ with mass ratio $\eta$ varying between $1:1$ and $1:8$. In Fig.~\ref{fig:PAcompare} we compare these estimates to our exact numerical result. At low frequencies, the NR estimate performs really well, agreeing with the exact result much better than should be expected from the estimated error and also outperforming the PN estimate. At higher frequencies, the NR estimate loses accuracy and systematically underestimates $\rho$, and the PN expression surprisingly gives a better approximation to the exact result. 

\begin{figure}[tb]
\includegraphics[width=\columnwidth]{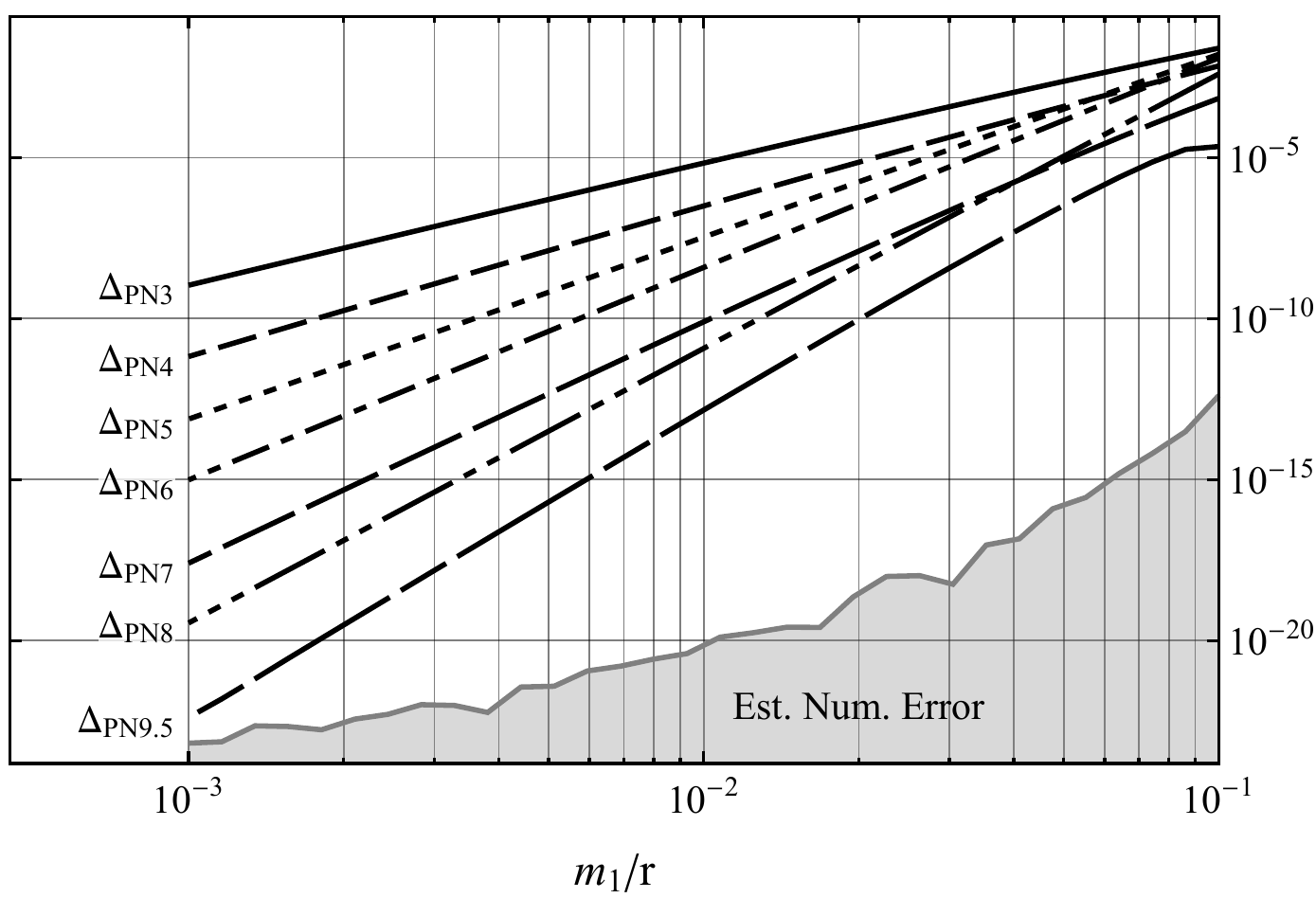}
\caption{Log-Log plot of the residual differences $\Delta_{\mathrm{PN}n}:= \abs{\rho_{\mathrm{GSF}}-\rho_{\mathrm{PN}n}}$ between our calculation of $\rho_{\mathrm{GSF}}$ at $a=0$ and successive PN approximants $\rho_{\mathrm{PN}n}$ provided in \cite{Bini:2016qtx}. The shaded area indicates the estimated numerical error on the self-force result.}
\label{fig:PNresiduals}
\end{figure}

In the non-spinning ($a=0$) case, much more accurate PN approximations for $\rho$ (up to 9.5PN) are available \cite{Bini:2016qtx}. To compare to these results we prepared a dense set of measurements of $\rho$ in the range $10m_1 <r <1000m_1$, accurate to 1 part in $10^{19}$ in the weak field. Figure~\ref{fig:PNresiduals} shows the residuals from subtracting successive PN approximants, $\Delta_{\mathrm{PN}n}:= \abs{\rho_{\mathrm{GSF}}-\rho_{\mathrm{PN}n}}$. In the weak field, we see a consistent improvement in the agreement as the PN order is increased, serving as a validation both of the high order PN approximants and of the high accuracy claimed for our results.

Recently, there have been two independent derivations \cite{Jaranowski:2013lca,Damour:2014jta,Jaranowski:2015lha,Bernard:2015njp} of the full 4PN equations of motion for non-spinning binaries. Their results agree on almost all coefficients of the PN expansion, except for a couple of linear in mass ratio terms, which crucially lead different contributions to the periapsis advance. The results of \cite{Bini:2016qtx} agree with \cite{Jaranowski:2013lca,Damour:2014jta,Jaranowski:2015lha}, but depend on filtering a PN expansion of the redshift invariant through the first law of binary mechanics and the EOB formalism. It is therefore of interest to provide a completely independent estimate of this coefficient that does not depend on any such theoretical bridge. We do so by fitting a PN series of the form
\begin{equation}
\begin{split}
\rho(x) = &\sum_{i=2}^\infty \rho_{ic} x^i+ \sum_{j=5}^\infty\rho_{jh} x^{j+1/2}
+\sum_{k=4}^\infty\rho_{kl} x^{k}\log(x)\\
&+\sum_{n=8}^\infty\rho_{nl2} x^{n}\log^2(x) + \dots
\end{split}
\end{equation}
to our dense data set. If we make no other assumptions than this functional form of the series we find for the 4PN non-log term $\rho_{4c} =64.5(1)$. The accuracy of this fit can be increased by assuming exact values for the known PN coefficients. Including values for the 3PN coefficients and the 4PN log term (that both calculations agree upon), we find $\rho_{4c} =64.64049(8)$. If we assume all known PN coefficients except $\rho_{4c}$, taking the values given in \cite{Bini:2016qtx},  the accuracy is further increased to $\rho_{4c} =64.640564757116(4)$. This is in perfect agreement to the exact value given in \cite{Bini:2016qtx} (and consequently \cite{Jaranowski:2013lca,Damour:2014jta,Jaranowski:2015lha}), which equates to $\rho_{4c} \approx 64.640564757119...$ .

\begin{figure}[tb]
\includegraphics[width=\columnwidth]{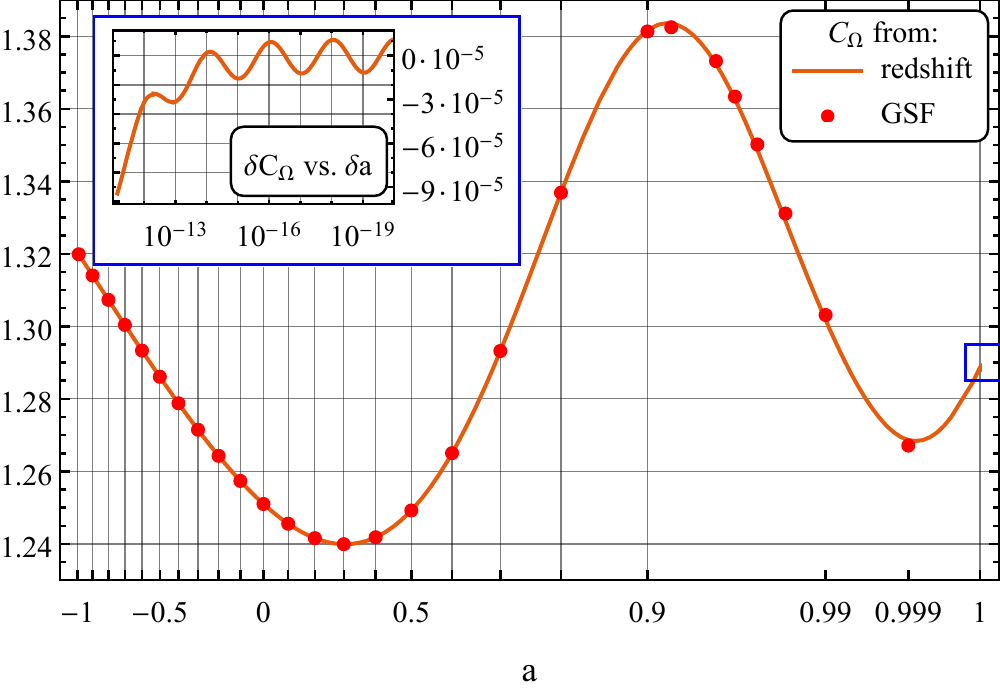}
\caption{Comparison of two methods for calculating the ISCO shift from either  GSF or redshift data. The x-axis has had a non-linear scaling applied to better display the new phenomenology in the $a>0.9$ region. The inset shows a close-up of the near extremal limit plotting $\delta{C}=C_\Omega-C_1$ vs. $\delta{a}=1-a$, revealing persistent order $10^{-5}$ oscillations.
}\label{fig:ISCO}
\end{figure}

We next calculate the ISCO shift. As mentioned above, we have two independent ways of calculating the shift of the ISCO; One using the GSF to calculate the periapsis advance, taking the limit towards the ISCO, and using Eq. \eqref{eq:GSFisco}. The other, described in \cite{Isoyama:2014mja}, using data for the redshift on circular orbits (which we calculate using the implementation of \cite{vandeMeent:2015lxa}). Fig.~\ref{fig:ISCO} plots the results of both calculations finding excellent agreement (5-6 digits, consistent with numerical error). This result bolsters the credence of the ingredients used in the method of \cite{Isoyama:2014mja}, including the first law of binary mechanics.

Figure~\ref{fig:ISCO} also extends the results of \cite{Isoyama:2014mja} (which covered spins $-0.9<a<0.9$) to much higher spins approaching extremality. This calculation was done solely using the redshift method, which is faster by at least an order of magnitude due to needing only data on circular orbits. The calculation is further aided by simplifications of the MST method in the near-horizon near-extremal Kerr (NHNEK) limit \cite{Meent:2015a}. This calculation reveals a much richer structure than implied by \cite{Isoyama:2014mja}. Shortly after $a=0.9$ the ISCO shift reaches a maximum, after which it decreases to another minimum to ultimately appear to monotonically approach a limit value, $C_1$. This limit value may be traced back to coming solely from mass and angular momentum perturbation contributions to the redshift. Consequently, it may be calculated analytically \cite{Warburton:2016unpublished} to obtain $C_1 = 1+1/(2\sqrt{3})$. However, on closer examination of the NHNEK limit we find that $C_\Omega$ continues to oscillate around this limit value with an amplitude of the order of $\sim 10^{-5}$. Similar oscillations as a function of $\delta{a}=1-a$ have previously been observed in calculations of other observable quantities in the NHNEK limit, including the quasinormal-mode frequencies\cite{Yang:2013uba}, and gravitational wave flux \cite{Gralla:2015rpa}. Yet, the author is unaware of any intuitive geometrical explanation of their probable common origin.

\prlsection{Discussion}
In this Letter, we have produced the first direct calculation of invariant observables (periapsis advance and ISCO shift) sensitive to the conservative part of the gravitational self force on eccentric orbits of spinning binaries. We expect these to be key benchmarks for the coming years in improving modelling for eccentric spinning binaries and, in particular, in the push for getting more faithful models at lower mass-ratios. This benchmark function has been demonstrated by discriminating between two competing derivations of the 4PN equations of motion for non-spinning binaries.

The calculation of the ISCO shift has been compared with earlier calculations \cite{Isoyama:2014mja} based on calculation of the redshift on circular orbits. The excellent agreement serves as a verification of some of the novel elements of our calculation such as the gauge completion of the radiation gauge results. Moreover, it provides a validation of the theoretical underpinnings of \cite{Isoyama:2014mja}. In particular, it validates the first law of binary mechanics in a regime where it has not been tested.

The examination of the ISCO shift in the near-extremal regime has revealed interesting new phenomenology. In particular the oscillation of the ISCO shift as the spin approaches extremality seems interesting, as it implies that the ISCO shift does not have a proper extremal limit. These oscillations beg for an explanation, either in terms of the extremal spacetime geometry, or in terms of a Kerr/CFT dual \cite{Guica:2008mu,Compere:2012jk}.

\begin{acknowledgments}
\textit{Acknowledgments.\textendash{}} The author thanks Alexandre Le Tiec for providing the error bars for Fig.~\ref{fig:PAcompare}. He also thanks Leor Barack for feedback on an early version of this manuscript. The author was supported by the European Research Council under the European Union's Seventh Framework Programme (FP7/2007-2013) ERC grant agreement no. 304978. The numerical results in this Letter were obtained using the IRIDIS High Performance Computing Facility at the University of Southampton.
\end{acknowledgments}

\raggedright
\bibliography{../bib/journalshortnames,../bib/meent,../bib/commongsf,PAshift}

\end{document}